\documentclass[a4paper]{jpconf}
\usepackage{graphicx}
\usepackage{bm}
\usepackage{amsmath}

\newcommand{\thcrsi}{ThCr$_{2}$Si$_{2}$}
\newcommand{\dyrusi}{DyRu$_{2}$Si$_{2}$}

\newcommand{\dyyrusi}{Dy$_{x}$Y$_{1-x}$Ru$_{2}$Si$_{2}$}
\newcommand{\femntio}{Fe$_{x}$Mn$_{1-x}$TiO$_{3}$}

\newcommand{\Tg}{T_{\mbox{\scriptsize g}}}

\newcommand{\Hdc}{H_{\mbox{\scriptsize dc}}}

\newcommand{\chieq}{\chi_{\mbox{\scriptsize eq}}}
\newcommand{\Meq}{M_{\mbox{\scriptsize eq}}}
\newcommand{\gammaMF}{\gamma_{\mbox{\scriptsize MF}}}
\newcommand{\zMF}{z_{\mbox{\scriptsize MF}}}
\newcommand{\nuMF}{\nu_{\mbox{\scriptsize MF}}}
\newcommand{\znuMF}{(z\nu)_{\mbox{\scriptsize MF}}}

\begin{document}
\title{Mean-field type Ising spin glass in a diluted rare-earth intermetallic compound}

\author{Y Tabata$^{1}$, K Matsuda$^{1}$, S Kanada$^{1}$, T Waki$^{1}$, H Nakamura$^{1}$, K Sato$^{2}$ and K Kindo$^{2}$}

\address{$^{1}$Department of Materials Science and Engineering, Kyoto University, Kyoto 606-8501, Japan}
\address{ $^{2}$The Institute for Solid State Physics, University of Tokyo, Kashiwa 277-8581, Japan}

\ead{y.tabata@ht4.ecs.kyoto-u.ac.jp}

\begin{abstract}
We have investigated a phase transition of a long-range RKKY Ising spin glass  \dyyrusi\ by means of the dc and the ac magnetization measurements.  The static nonlinear susceptibility $\chi_{2}$ exhibits a divergent behavior toward $\Tg$ $=$ 1.925 K with the exponent of $\gamma$ $=$ 1.08.  The characteristic relaxation time $\tau$ obtained from the ac magnetization measurements also diverges to the same $\Tg$ with the exponent of $z\nu$ $=$ 1.9 in zero static field . These results indicate the spin glass phase transition belonging to the mean-field universality class, in which the exponents of $\gammaMF$ and $\znuMF$ are 1 and 2 respectively. Moreover, we observed the divergence of $\tau$ in a finite field of 500 Oe toward $\Tg$ $=$ 1.412 K with the exponent of $z\nu$ $=$ 2.1, being quite similar to that in zero field. It strongly suggests an existence of the spin glass phase transition in a finite field and the replica symmetry breaking (RSB) in the long-range Ising spin glass \dyyrusi .
\end{abstract}

In spite of the long-period investigation on the spin glass for more than 30 years, there are still several unsolved problems \cite{APYoung01,APYoung02}. One of them is whether the replica symmetry breaking (RSB) predicted by the mean-field theory of the spin glass \cite{Parisi} occurs or not in real spin glass materials. In the mean field picture, the spin glass state has a complicated multi-valley structure of the free-energy and survives in a finite magnetic field \cite{AT}. Alternative theory of the spin glass, so called the droplet theory based on the renormalization group arguments, predicts no RSB and that only two thermodynamic states related each other by a global spin flip exist in the spin glass state \cite{FisherandHuse}. In the droplet picture, no spin glass phase transition exists in the presence of a magnetic field. Many numerical studies have been employed extensively on a stability of the spin glass phase in the presence of magnetic field \cite{Numerical01,Numerical02,Numerical03,Numerical04,Numerical05} , however it is still controversial.  Recent experimental study of the short-range Ising spin glass  \femntio\ gave an evidence against an equilibrium phase transition in a finite magnetic field and indicated that the droplet picture is appropriate \cite{Nordblad, Petra}. On the other hand, the magnetic torque measurements of several Heisenberg spin glasses exhibit the existence of true phase transitions in fields \cite{Petit}. 

In this article, we reported recent ac and dc magnetization measurements for the Ising spin glass \dyyrusi\ . \dyrusi\ is a rare-earth intermetallic compound with the tetragonal \thcrsi -type crystal structure, exhibiting two successive antiferromagnetic transitions with $T_{\mbox{\scriptsize N}1}$ $=$ 29 K and $T_{\mbox{\scriptsize N}2}$ $=$ 3.5 K \cite{Iwata}. The compound has a strong uniaxial magnetic anisotropy and recognized as an Ising antiferromagnet. In the diluted system \dyyrusi , where the magnetic Dy$^{3+}$-ion is substituted by the non-magntic Y$^{3+}$-ion, a spin glass transition is undergone with $x$ $=$ 0.103. In this compound, the Ising Dy$^{3+}$-moments interact via the long-range RKKY interaction, and hence, its spin glass phase transition could belong to a different universality class from the short-range Ising spin glass such as \femntio . Our analyses of the static and the dynamic critical phenomena present that \dyyrusi\ exhibits the mean field type phase transition in zero and finite fields where the replica symmetry should be broken.

The single crystalline sample of \dyyrusi\ with $x$ $=$ 0.103 was grown by the Czhochralski method with a tetra-arc furnace. The concentration of Dy was determined by comparing the saturated magnetization of the diluted compound along the magnetic easy c-axis with that of the pure compund \dyrusi . The ac and the dc magnetizations were measured  with use of a SQUID magnetometer MPMS (Quantum Design). The ac measurements were performed with an ac-field of 3 Oe and frequency of 0.5 Hz $\leq$ $\omega$ $\leq$ 1000 Hz.

\begin{figure}[h]
\includegraphics[width=34pc]{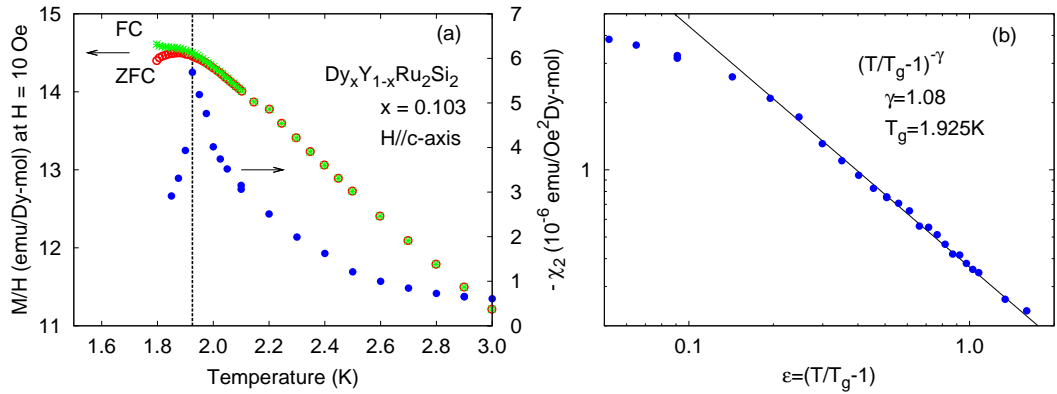}%
\caption{\label{fig1} (a) Temperature dependences of the dc magnetizations divided by the magnetic field $M/H$ at $H$ $=$ 10 Oe with the ZFC and the FC conditions. The temperature dependence of the negative nonlinear susceptibility $-\chi _{2}$ is also shown. The vertical dashed line represents the spin glass transition temperature $\Tg$. (b) Log-Log plot of $-\chi _{2}$ vs. the reduced temperature $\varepsilon$ $=$ $(T/\Tg -1)$.}
\end{figure}

Figure \ref{fig1} (a) shows the temperature dependence of the dc magnetization divided by the magnetic field $M/H$ at $H$ $=$ 10 Oe. The dc magnetization was measured with the zero-field-cooled (ZFC) and the field-cooled (FC) conditions. A distinct deviation of the ZFC and the FC magnetizations was found around 1.9 K, which is a characteristic of a spin glass phase transition. In Fig.\ref{fig1} (a) the temperature dependence of the negative nonlinear susceptibility $-\chi _{2}$ is also shown, which is the order parameter susceptibility of the spin glass phase transition in zero field limit. The nonlinear susceptibility $\chi_{2}$ was obtained by fitting the magnetization divided by the magnetic field $M(H,T)/H$ at each temperature as a function of $H^{2}$ in the form of, 
\[
M(H,T)/H = \chi_{0}(T) + \chi_{2}(T) H^{2} + \chi_{4}(T) H^{4} + \dotsb .
\]
As shown in Fig. \ref{fig1} (a), $\chi_{2}$ exhibits negative divergent behavior toward $\sim$ 1.9 K. To examine the critical divergence of $\chi _{2}$, a log-log plot of $-\chi_{2}$ against the reduced temperature $\varepsilon$ $\equiv$ $(T/\Tg -1)$ is shown in Fig. \ref{fig1} (b). The best plot showing a linear relation between $\log (-\chi_{2})$ and $\log \varepsilon$ is obtained with $\Tg$ $=$ 1.925 K and the critical exponent $\gamma$ $=$ 1.08.  The value of $\gamma$ is very close to $\gammaMF$ $=$ 1 predicted in the mean-field theory \cite{APYoung01}. This strongly suggests that the spin glass phase transition in \dyyrusi\ is the mean-field type. 

\begin{figure}[h]
\includegraphics[width=34pc]{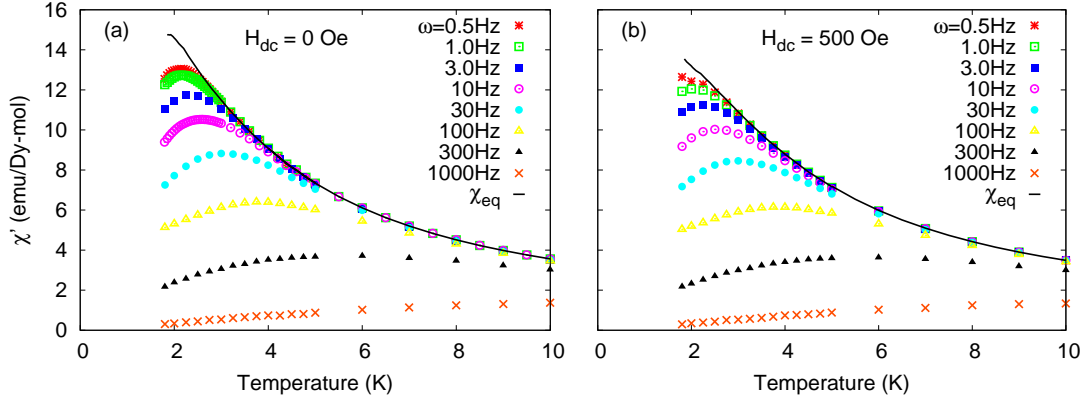}%
\caption{\label{fig2} Temperature dependences of the real part of the ac susceptibility $\chi '$ with $\Hdc$ $=$ (a) 0 Oe and (b) 500 Oe. In both (a) and (b), the solid lines represent the thermal equilibrium susceptibilities $\chieq$ at respective dc fields.} 
\end{figure}

Figure \ref{fig2} (a) and (b) show the temperature dependences of the real part of the ac susceptibility $\chi'$ with zero bias dc field and a finite field of 500 Oe, respectively. In the figures, thermodynamic equilibrium susceptibilities $\chieq$ at respective fields are also shown, which is determined as $\chieq (T,H)$ $=$ $\mbox{d}\Meq (T,H)/\mbox{d}H \mid _{H = \Hdc}$ from the dc magnetization measurements. In both zero and finite fields, the ac susceptibilities deviate from the equilibrium ones in low temperature regions and show broad maximums. The temperature showing the maximum increases with increasing the frequency. 
%The maximum of $\chi '$ represents a spin freezing in a finite observation time ($=$ $1/\omega$), which does not represent a thermal equilibrium phase transition. For evidence it from the dynamical aspect, they are necessary to determine the characteristic relaxation time $\tau$ and to observe its critical divergence. 
The maximum of $\chi '$ does not represent a thermal equilibrium phase transition but a spin freezing in a finite observation time ($=$ $1/\omega$). According to the dynamic scaling hypothesis \cite{Hohen}, the characteristic relaxation time $\tau$ exhibits the critical divergence as $\tau \propto \varepsilon ^{-z\nu}$, where $z$ and $\nu$ are the dynamic and the correlation length critical exponents respectively.  In this study, we identify $\tau (T,H)$ ($=$ $1/\omega$) in the criterion of $\chi' (\omega;T,H)$ $=$ $0.9 \chieq (T,H)$. A similar criterion was also used to determine $\tau$ in Ref. \cite{Petra}. 

\begin{figure}[h]
\includegraphics[width=34pc]{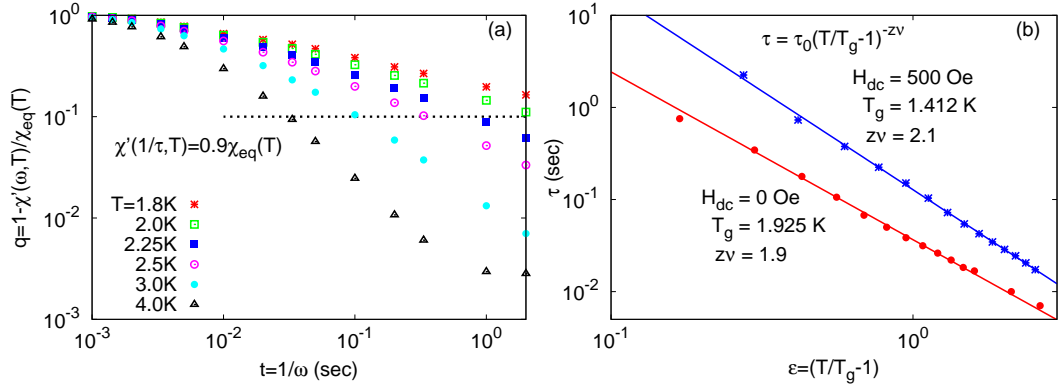}%
\caption{\label{fig3} (a) Dynamic spin correlation $q(t)$ $=$ $1-\chi' (\omega, T)/\chieq(T)$ with $t$ $=$ $1/\omega$ in zero field as a function of time on a log-log scale. (b) Log-log plots of  the characteristic relaxation time $\tau$ vs. $\varepsilon$ $=$ $(T/\Tg -1)$ at $H$ $=$ 0 and 500 Oe.}
\end{figure}

We plot $q(t)$ $=$ $1-\chi' (\omega, T)/\chieq(T)$ with $t$ $=$ $1/\omega$, corresponding the dynamic spin correlation function, in zero field against time at several temperatures in Fig. \ref{fig3} (a).  %%%
Above criterion to identify $\tau$, $q(t)$ $=$ $0.1$, is represented by the horizontal dashed line in the figure. $\tau (T,H)$ at $H$ $=$ 500 Oe was also determined by using the same plot. %%%
Because the frequency dependence of the ac susceptibility of \dyyrusi\ is very strong, as shown in Fig. \ref{fig2}, we can see the decay of the spin correlations over three decades in magnitude at 4 K in Fig. \ref{fig3} (a). As decreasing temperature, the dynamics becomes slower, and $q(t)$ does not reach 0.1 below 2 K, which is very close to $\Tg$ $=$ 1.925 K obtained from the static critical scaling analysis of the nonlinear susceptibility. It is consistent with the general feature of $\tau$, being infinite below $\Tg$, and hence, we concluded that our criterion to identify $\tau$ is appropriate for the present compound. 

Figure \ref{fig3} (b) shows log-log plots of $\tau (T,H)$ vs. $\varepsilon$ $=$ $(T/\Tg -1)$ at $H$ $=$ 0 and 500 Oe, respectively. In zero field, we assigned $\Tg$ $=$ 1.925 K, being the same value obtained from the static critical scaling analysis of the nonlinear susceptibility, and obtained the dynamic critical exponent $z\nu$ $=$ 1.9. Moreover, the analysis of $\tau (T,H)$ at $H$ $=$ 500 Oe gives a finite spin glass transition temperature $\Tg(H)$ $=$ 1.412 K and quite similar critical exponent $z\nu$ $=$ 2.1 to that in zero field. These results strongly suggest a presence of the thermal equilibrium spin glass phase even in a finite magnetic field in \dyyrusi . The value of $z\nu$ is much smaller than that in other spin glasses, for instance $z\nu$ $\approx$ 11 in \femntio\ \cite{Nordblad,Petra}. On the other hand, it is consistent with the prediction of the mean-field theory, where $\znuMF$ is 2 with $\zMF$ $=$ 4 and $\nuMF$ $=$ 1/2 \cite{APYoung01}.

In conclusion, the static scaling analysis of the nonlinear susceptibility and the dynamic scaling analysis of the characteristic relaxation time reveal the presence of the mean-field type phase transition in the long-range RKKY Ising spin glass \dyyrusi\ in both zero and finite magnetic fields. The validity of the mean-field picture indicates that the replica symmetry is broken in the spin glass state of the compound.

\end{document}